\documentclass[9pt,conference]{IEEEtran}
\IEEEoverridecommandlockouts

\usepackage{pgfplots}
\pgfplotsset{compat=1.18}
\usepackage{pgfplotstable}
\usetikzlibrary{patterns}

\usepackage{cite}
\usepackage{amsmath,amssymb,amsfonts}
\usepackage{algorithmic}
\usepackage{graphicx}
\usepackage{textcomp}
\usepackage{xcolor}
\usepackage{url}
\usepackage{algorithmic}
\usepackage{algorithm}
\usepackage{multirow}
\usepackage{hyperref}
\usepackage{fancyhdr}

\def\BibTeX{{\rm B\kern-.05em{\sc i\kern-.025em b}\kern-.08em
    T\kern-.1667em\lower.7ex\hbox{E}\kern-.125emX}}

\definecolor{mypurple}{HTML}{5E1675}
\definecolor{mygreen}{HTML}{337357}
\definecolor{myyellow}{HTML}{FA9E05}
\definecolor{myblue}{HTML}{284E78}
\definecolor{myred}{HTML}{E7324C}

\begin{document}

\title{Self-Tuning Spectral Clustering for Speaker Diarization}

\author{\IEEEauthorblockN{Nikhil Raghav\textsuperscript{1,3}, Avisek Gupta\textsuperscript{1}, Md Sahidullah\textsuperscript{1,2}, Swagatam Das\textsuperscript{1,2,4}}
\IEEEauthorblockA{$^1$Institute for Advancing Intelligence, TCG CREST, Kolkata-700 091, India\\
  $^2$Academy of Scientific and Innovative Research (AcSIR), Ghaziabad-201 002, India \\
  $^3$Department of Computer Science, RKMVERI, Howrah-711 202, India \\ $^4$ Electronics and Communication Sciences Unit, Indian Statistical Institute, Kolkata-700 108, India\\
e-mail: \{nikhil.raghav.92, md.sahidullah,  avisek.gupta, swagatam.das\}@tcgcrest.org}}

\maketitle

\begin{abstract}
Spectral clustering has proven effective in grouping speech representations for speaker diarization tasks, although post-processing the affinity matrix remains difficult due to the need for careful tuning before constructing the Laplacian. In this study, we present a novel pruning algorithm to create a sparse affinity matrix called \emph{spectral clustering on p-neighborhood retained affinity matrix} (SC-pNA). Our method improves on node-specific fixed neighbor selection by allowing a variable number of neighbors, eliminating the need for external tuning data as the pruning parameters are derived directly from the affinity matrix. SC-pNA does so by identifying two clusters in every row of the initial affinity matrix, and retains only the top $p\%$ similarity scores from the cluster containing larger similarities. Spectral clustering is performed subsequently, with the number of clusters determined as the maximum eigengap. Experimental results on the challenging DIHARD-III dataset highlight the superiority of SC-pNA, which is also computationally more efficient than existing auto-tuning approaches. Our implementations are available at \href{https://github.com/nikhilraghav29/SC-pNA}{https://github.com/nikhilraghav29/SC-pNA}.

\end{abstract}

\thispagestyle{fancy}
\fancyhf{}
\renewcommand{\headrulewidth}{0pt}
\chead{\small Accepted to be Published in 2025 International Conference on Acoustics, Speech, and Signal Processing, 2025, Hyderabad, India}
\pagestyle{empty}
\lfoot{\small \copyright 2025 IEEE. Personal use of this material is permitted. Permission from IEEE must be obtained for all other uses, in any current or future media, including reprinting/republishing this material for advertising or promotional purposes, creating new collective works, for resale or redistribution to servers or lists, or reuse of any copyrighted component of this work in other works.}


\begin{IEEEkeywords}
speaker diarization, spectral clustering, matrix sparsification, eigengap, DIHARD-III
\end{IEEEkeywords}

\section{Introduction}
The availability of precise annotations of audio recordings based on speaker information can significantly enhance the field of audio analytics involving multi-speaker conversations. This has applications across various domains, from everyday online meetings to advanced speech and audio forensics. This task is formally known as \emph{speaker diarization} (SD), where the goal is to assign speaker labels to segments of speech based on speaker identity~\cite{park2022review}. Traditionally, SD addresses the problem of ``who spoke when'' in a given audio recording, particularly in multi-speaker environments. The SD pipeline conventionally consists of several interconnected components, including \emph{speech enhancement} (SE), \emph{speech activity detection} (SAD), \emph{segmentation}, \emph{speaker embedding extraction}, \emph{clustering}, and \emph{re-segmentation}. 

Modern SD architectures based on deep learning~\cite{dawalatabad21_interspeech} have demonstrated remarkable performance, surpassing classical methods on widely used benchmark datasets such as CALLHOME~\cite{CallHome}, AMI~\cite{carletta2005ami}, and VoxSRC~\cite{huh2023voxsrc}. However, SD systems still face practical challenges when applied to more realistic conversational speech datasets, such as the DIHARD-III dataset~\cite{ryant21_interspeech}. Factors like background noise, overlapping speech, a large number of speakers, imbalanced speech contributions across speakers, intra-speaker variability, and variability in the recording environments make the task even more complex.
While issues such as environmental noise, SAD, speech overlap, and intra-speaker variability are extensively studied, the clustering problem does not receive the necessary attention despite playing a crucial role in the overall diarization process.

Clustering is a classical problem in unsupervised machine learning with a substantial body of literature~\cite{jain2010data,ren2024deep}. However, in the context of SD, clustering algorithms are largely limited to basic techniques such as \emph{k-means}, \emph{agglomerative hierarchical clustering}, \emph{mean-shift}~\cite{salmun2017plda}, and \emph{spectral clustering} (SC)~\cite{ning2006spectral}. Among these, SC is the most widely used, particularly in modern SD systems based on deep speaker embeddings~\cite{Park2020AutoTuningSC}. This preference is primarily due to its simplicity, with fewer parameters, and its strong mathematical foundations~\cite{ng2001spectral,von2007tutorial}.

The spectral clustering approach used for the SD task involves specific parameters during the adjustment of the affinity matrix. Since the nodes represented by speaker embeddings capture speaker characteristics, it is crucial to minimize the impact of different speaker similarities and to retain or enhance same-speaker similarities~\cite{Park2020AutoTuningSC}. In practice, a sparse graph is formed by removing less similar nodes and setting their values to zero. This process requires a pruning parameter typically tuned using a development set containing labeled speech data. However, audio from different domains may introduce domain mismatches, substantially degrading SD performance. Recently, an auto-tuning approach for SC, named as \emph{auto-tuning spectral clustering} (ASC), has been proposed in the context of SD, eliminating the need for tuning on external data~\cite{Park2020AutoTuningSC}. The tuning parameters are instead derived directly from the audio recording being evaluated, using a brute-force search across different parameter values and optimizing a proxy evaluation metric. While this technique is promising, it has several limitations. First, it requires repeated spectral clustering computations involving eigenvalue decomposition, making it computationally expensive for longer recordings. Second, the proxy evaluation metric has limitations and does not always accurately reflect actual diarization performance~\cite{Park2020AutoTuningSC}. Third, it uses a fixed number of neighbors for each node, which is problematic, especially for audio recordings with an imbalanced distribution of speech. Additionally, the graph may become disconnected in specific situations, leading to the failure of the spectral clustering algorithm, or may otherwise produce suboptimal results~\cite{von2007tutorial,shi2000normalized}.

This work introduces a novel self-tuning method called spectral clustering on p-neighborhood retained affinity matrix
(SC-pNA) for SD, addressing the limitations of previous state-of-the-art (SOTA) adaptive spectral clustering approaches. Inspired by the score distribution characteristics reported in speaker recognition literature~\cite{poh2005f,prakash07_interspeech}, we propose a score-based, node-specific pruning strategy. In this method, the pruning threshold is dynamically adjusted for each node, allowing the retention of nodes with stronger connections while eliminating weakly connected, unreliable nodes. Like ASC, our approach does not require external data for threshold adjustment. However, it advances further by eliminating the need for proxy evaluation metrics. Additionally, it requires only a single eigenvalue decomposition of the Laplacian matrix, which is computed once from the refined adjacency matrix.


\section{Spectral clustering for speaker diarization}

\subsection{Conventional spectral clustering (CSC)}
Spectral clustering ~\cite{von2007tutorial, shi2000normalized} is a technique used to group data points into clusters based on the eigenvalues and eigenvectors of a similarity matrix constructed from the data. It leverages the properties of the graph Laplacian to partition the data in a way that often captures complex cluster structures better than traditional methods like $k$-means. With the emergence of speech embeddings~\cite{dehak2010front,snyder2018x,bai2021speaker}, SC is widely adopted for SD~\cite{ning2006spectral,shum12_interspeech,lin19_interspeech,dawalatabad21_interspeech}. 

The semi-supervised CSC~\cite{dawalatabad21_interspeech} for SD aims to identify $k$ speaker classes from a finite set 
$\mathbf{X} = \{ \mathbf{x}_1, \mathbf{x}_2, \dots, \mathbf{x}_n \}$ of $n$ embeddings,
extracted from $n$ speech segments. CSC first computes an affinity matrix 
$\mathbf{M} \in \mathbb{R}^{n \times n}$,
containing the pairwise cosine similarities between speech embeddings. 
Next, $\left \lfloor n(1-\alpha) \right \rfloor$ similarity scores are deemed unreliable and hence pruned in every row of $\mathbf{M}$, using a single pruning parameter $\alpha$.
The resulting matrix $\mathbf{M}_{\alpha}$ is symmetrized, and an unnormalized Laplacian matrix is computed as $\mathbf{L} = \mathbf{D} - \mathbf{M}_{\alpha}$, where $\mathbf{D}$ is a degree matrix. 
The number of speakers $k$ is estimated using the eigengap approach \cite{von2007tutorial}, and finally, the $k$-means algorithm clusters the $k$ eigenvectors of $\mathbf{L}$ corresponding to its $k$ smallest eigenvalues.
CSC is semi-supervised, as the optimal pruning parameter $\alpha$ exhibiting minimum \emph{diarization error rate} (DER) is estimated from a development set, by varying $\alpha$ linearly from 0 to 1 with a typical step size of 0.01. 
This setting is however not the best choice when the development and evaluation data comes from substantially different environments~\cite{raghav2024assessing}. 

\subsection{Auto-tuning spectral clustering (ASC)}
To not depend on a development set for parameter tuning, an unsupervised variant of SC known as \emph{auto-tuning spectral clustering} (ASC) was proposed in~\cite{Park2020AutoTuningSC}. 
ASC estimates a proxy DER from the unlabelled data. 
This is done by first determining the maximum eigengap, normalized by the largest eigenvalue, and then computing the ratio of the pruning factor to the maximum eigengap. 
The proxy DER however may not always be a good approximation of the original DER. 
Additional limitations of both ASC and CSC are their high computation costs. This is due to the repeated eigendecomposition of the Laplacian formed at varying pruning rates, which is required to estimate the maximum eigengap.
Both methods also prune a strictly fixed number of entries across all rows of the affinity matrix, which is also not an ideal choice in situations when the distribution of spoken duration of speakers is not uniform across a recording.

\subsection{Spectral clustering on $p$-neighborhood retained affinity matrix (SC-pNA)}
\textbf{Background:} Our proposed method of SC-pNA is focused on creating a sparse affinity matrix that can aid the spectral clustering method to better delineate different speaker embeddings, thereby improving SD. To create this sparse affinity matrix, each row is pruned independently by removing low similarity scores. Each row of the original affinity matrix should ideally have high similarities to the embeddings of the same speakers, while having low similarities to the embeddings of all other speakers.   

As the number of embeddings belonging to the same speaker will vary across a recording, therefore the number of similarity scores retained in each row should be decided adaptively. In general, the similarity scores in each row of an affinity matrix can be described in terms of two distributions. The first \textit{within-cluster} distribution $C_w$ corresponds to the high similarity scores between the same speaker embeddings, whereas the second \textit{between-cluster} distribution $C_b$ describes the low similarity scores between different speaker embeddings. This approach of describing similarities in terms of two distributions has been previously explored in the context of speaker recognition~\cite{poh2005f,brummer2014generative}, where $C_w$ and $C_b$ are assumed to follow the Gaussian distribution, i.e., $C_w \sim \mathcal{N}(\mu_w, \sigma_w)$, and $C_b \sim \mathcal{N}(\mu_b, \sigma_b)$. As we wish to accept $C_w$ and reject $C_b$, the \emph{equal error rate} (EER) provides a location where the false rejection rate of $C_w$ equals the false acceptance rate of $C_b$. 
The analytical EER is defined as, 
\begin{equation}
    \text{EER} = \frac{1}{2} - \frac{1}{2}\text{erf}\left(\frac{\text{F-ratio}}{\sqrt{2}}\right),
\end{equation}
where erf(.) is the error function for Gaussian distributions, and the F-ratio is $(\mu^w - \mu^b)/(\sigma^w + \sigma^b)$. 
The optimal threshold corresponding to the EER, shown in Figure \ref{fig:gaussians}, is given by,
\begin{equation} \label{eq:delta}
    \Delta = \frac{\mu^w \sigma^b + \mu^b \sigma^w}{\sigma^w + \sigma^b}.
\end{equation}

The theoretical foundations of the threshold $\Delta$ make it an appealing choice to delineate distributions $C_w$ and $C_b$ in each row of the affinity matrix, and thereafter all entries corresponding to $C_b$ can be set to zero to create the sparse affinity matrix. 
We refer to this method as the EER-$\Delta$ approach: on an initial affinity matrix $\mathbf{A}$, $\Delta_i$ is first computed over each row $\mathbf{A}_i$, for $i=1,...,n$. 
A sparse affinity matrix $\mathbf{B}$ is then created by using $\Delta_i$ as a threshold in each row $\mathbf{B}_i$, where we set $\mathbf{B}_{ij} = 0$ if $\mathbf{A}_{ij}<\Delta_i$, otherwise $\mathbf{B}_{ij} = \mathbf{A}_{ij}$. Finally, spectral clustering is conducted on $\mathbf{B}$, using the maximum eigengap to determine the number of clusters. 
The EER-$\Delta$ approach has a lower computation cost compared to the CSC and ASC, as it only computes the eigendecomposition of a single Laplacian, formed from the affinity matrix whose rows are pruned using $\Delta_i$.
An additional advantage of EER-$\Delta$ is that the number of similarity scores pruned in each row is adaptive, and depends on the sizes of the identified $C_w$ and $C_b$.

\begin{figure}[!ht]
    \centering
    \begin{tikzpicture}[thick,scale=0.6, every node/.style={scale=0.6}]
    \begin{axis}[
        no markers, 
        domain=-9:12, 
        samples=200, 
        axis lines*=left, 
        xlabel=$\boldsymbol{x}$,
        ylabel=$\boldsymbol{f(x)}$,
        height=7cm, 
        width=16cm,
        enlargelimits=false, 
        clip=false, 
        grid = major,
        grid style={dashed, gray!30},
        legend style={at={(0.5,-0.15)}, anchor=north,legend columns=-1},
        ytick=\empty,
        xtick={-8.8, 11.8},
        xticklabels={-1, 1},
        tick label style={font=\Large},
        axis line style={thick},
        xlabel style={font=\Huge},
        ylabel style={font=\Huge}
    ]
    \addplot [line width=1.5pt,color=mygreen] {0.98*exp(-0.2*((x+2)^2)/1.4)};
    \node at (axis cs:-2,0.8) {\Huge $\mathbf{C_b}$};

    \addplot [line width=1.5pt,color=mypurple] {1.1*exp(-0.2*((x-5)^2)/0.7)};
    \node at (axis cs:5,0.93) {\Huge $\mathbf{C_w}$};
    \draw[line width=3pt,dashed,color=myred] (axis cs:2.06,0) -- (axis cs:2.06,0.19);

    \node at (axis cs:2.06,0.51) {\Huge EER};
    \node at (axis cs:1.82,0.38) {\Huge optimal};
    \node at (axis cs:2.06,0.25) {\Huge $\Delta$};
    
    \draw[line width=3pt,dashed,color=myyellow] (axis cs:6.5,0) -- (axis cs:6.5,0.56);
    \addplot [
        domain=6.5:12, 
        samples=200, 
        pattern=north east lines, 
        pattern color=myyellow,
        draw=none,
    ] {1.1*exp(-0.2*((x-5)^2)/0.7)} \closedcycle;

    \node at (axis cs:7.8,0.5) {\Huge $\mathbf{p}\%$};
    \node at (axis cs:8.6,0.35) {\Huge retention};
    \end{axis}
    \end{tikzpicture}
    \caption{The EER-$\Delta$ approach accepts all similarity scores in cluster $C_w$ up to the EER optimal $\Delta$ threshold, whereas the proposed SC-pNA only retains the top $p\%$ similarity scores in cluster $C_w$, and prunes all lower similarities.}
    \label{fig:gaussians}
\end{figure}

\textbf{SC-pNA Methodology:} In SC-pNA, we consider an even more aggressive thresholding approach, to create a sparser affinity matrix. As the $\Delta_i$ thresholds of EER-$\Delta$ retain all similarity scores for $C_w$ in the $i$-th row, we can interpret this approach as restricting the neighborhood of the $i$-th speaker segment $\mathbf{x}_i$ only to the $C_w$ cluster. Let $N(\mathbf{x}_i)$ be the set of speaker segments in the neighborhood of $\mathbf{x}_i$, and let $C_w^{(i)}$ be the cluster identified for $\mathbf{x}_i$. Then each $\Delta_i$ threshold of EER-$\Delta$ establishes, 
\begin{equation} \label{eq:EER}
    N(\mathbf{x}_i) = C_w^{(i)} \text{ for }  i = 1, 2, ..., n,
\end{equation}
 throughout the affinity matrix.

In the proposed SC-pNA, a more aggressive pruning is pursued, where only the top $p\%$ similarities from the identified $C_w^{(i)}\backslash\{\mathbf{x}_i\}$ are retained in every row. This is shown in Figure \ref{fig:gaussians}, where the proposed pruning approach can be interpreted as considering higher false rejection ratio (FRR), with the thresholds empirically set at (1-p)\% FRR. The constant high self-similarity scores $s(\mathbf{x}_i,\mathbf{x}_i) = 1 $ are excluded from the process to identify the clusters $C_w^{(i)}$ and $C_b^{(i)}$, as they can behave as outliers when all other similarities $s(\mathbf{x}_i,\mathbf{x}_j), \forall j \neq i$, are much smaller than $s(\mathbf{x}_i,\mathbf{x}_i)$. This exclusion is interpreted as the removal of all self-loops, thus leading to an adjacency matrix where connectivity is maintained only through neighborhood connections. Thus retaining only top $p\%$ similarities creates the neighborhoods for all $\mathbf{x}_i$, 
\begin{equation} \label{eq:Prop}
    N_p(\mathbf{x}_i) \subseteq C_w^{(i)}\backslash\{\mathbf{x}_i\}. 
\end{equation}
Comparing equations \eqref{eq:EER} and \eqref{eq:Prop} lets us determine that $|N_p(\mathbf{x}_i)| \leq |N(\mathbf{x}_i)|$, and thus smaller neighborhoods are formed, leading to more sparse affinity matrices.
However, these smaller neighborhoods should be capable of identifying the correct speaker clusters. If a recording contains $k$ speakers, the embeddings should be clustered correctly into $k$ clusters $C_1, \ldots, C_k$, where each cluster $C_j = \{ \mathbf{x}_{j1}, \mathbf{x}_{j2}, \ldots, \mathbf{x}_{j|C_j|}\}$ contains $|C_j|$ embeddings. Thus $p$ should be chosen so that $|N_p(\mathbf{x}_i)|$ is small, yet cluster $C_j$ can be recovered by examining the neighborhoods of $\{ \mathbf{x}_{j1}, \mathbf{x}_{j2}, \ldots, \mathbf{x}_{j|C_j|}\}$, i.e.,
\begin{equation}
   N_p(\mathbf{x}_{j1}) \cup N_p(\mathbf{x}_{j2}) \cup \ldots \cup N_p(\mathbf{x}_{j|C_j|}) = C_j.
\end{equation}
In the EER-$\Delta$ approach, $N(\mathbf{x}_{j1}) \cup \ldots \cup N(\mathbf{x}_{j|C_j|}) = C_j$ holds trivially. For real world data, that can be noisy and have overlapped clusters,  the challenge for both SC-pNA and the EER-$\Delta$ approach lies in utilising the available neighborhood information to identify the set of embeddings 
$\{ \mathbf{x}_{j1}, \mathbf{x}_{j2}, \dots, \mathbf{x}_{j|C_j|} \}$ that recovers each cluster $C_j$ accurately.

The following are the steps of the proposed SC-pNA approach:
\begin{enumerate}
    \item \textbf{Create the adjacency matrix without self-loops:}
    The initial adjacency matrix $\mathbf{A}$ is formed using cosine similarity,
\begin{equation} \label{eq:Affinity}
    \mathbf{A}_{ij} = \frac{\mathbf{x}_i^\top \mathbf{x}_j}{\|\mathbf{x}_i\| \|\mathbf{x}_j\|}, \quad \forall i, j \in \{1, 2, \ldots, n\}, \ i \neq j.
\end{equation}

The diagonal entries of $\mathbf{A}$ are set to zero i.e.,  $\mathbf{A}_{ii} = 0, \ i = 1, 2, \ldots, n$.  
    \item \textbf{Identify $C_w^{(i)} \ \text{and} \ C_b^{(i)}$ in each row:}
    In each row $\mathbf{A}_i$, two clusters $C_1^{(i)}, C_2^{(i)}$ are identified using $k$-Means clustering with $k = 2$. From these two clusters, $C_w^{(i)}$ is identified as the one with the larger cluster center, 
    \begin{equation}\label{eq:Identify_clusters}
        C_w^{(i)} = \text{argmax}\{\text{center}(C_1^{(i)}), \text{center}(C_2^{(i)})\}.
    \end{equation}
    $ C_b^{(i)}$ is identified as the other cluster i.e., $C_b^{(i)} = \text{argmin}\{\text{center}(C_1^{(i)}), \text{center}(C_2^{(i)})\}$.
    \item \textbf{Retain top $p\%$ similarity scores:} 
    A matrix $\mathbf{P}$ is formed which retains in each row the top $p\%$ similarity scores from the cluster $ C_w^{(i)}$, and the rest are set to zero, i.e.,
    \begin{equation}\label{eq:pruning}
        P_{ij} = \begin{cases}
            0 & \text{, if } \mathbf{A}_{ij} \text{ is not in the top } p \% \text{ of } C_w^{(i)} \\ 
            \mathbf{A}_{ij} & \text{, otherwise}
        \end{cases}.
    \end{equation}
    \item \textbf{Symmetrize and form the Laplacian:}
    A symmetric matrix $\mathbf{W}$ is formed from $\mathbf{P}$, 
    \begin{equation}\label{eq:Symmetrize}
    \mathbf{W} = \frac{1}{2} \left( \mathbf{P} + \mathbf{P}^\top \right).
    \end{equation}

    A diagonal matrix $\mathbf{D}$ is created with the sum of the rows of $\mathbf{W}$ as its entries: $\mathbf{D}_{ii} = \sum_{j = 1}^n|w_{ij}|$.
    The Laplacian $\mathbf{L}$ is then created,
    \begin{equation}\label{eq:Laplacian}
        \mathbf{L} = \mathbf{D} - \mathbf{W}.
    \end{equation}
    \item \textbf{Compute eigengap and estimate $\hat{k}$ :}
    For each recording, a user can specify a reasonable assumption for the  maximum number of speakers $k_{\max}$. We find the smallest $M$ eigenvalues $\lambda_1, \ldots, \lambda_M$ of $\mathbf{L}$, where $M = \min\{k_{\max}, n\}$. The $M - 1$ length eigengap vector is constructed as, 
    \begin{equation}\label{eq:Eigengap}
        \mathbf{e_{gap}} = [\lambda_2 - \lambda_1, \lambda_3 - \lambda_2, \dots, \lambda_M - \lambda_{M-1}].
    \end{equation}
    The number of speakers $\hat{k}$ is then estimated as, 
    \begin{equation}\label{eq:Estimate_k}
        \hat{k} = \text{argmax} \ \mathbf{e_{gap}}. 
    \end{equation}
    \item \textbf{Cluster the spectral embeddings:}
    Form matrix $\mathbf{V} \in \mathbb{R}^{n \times \hat{k}}$, containing in its columns the $\hat{k}$ eigenvectors corresponding to the smallest $\hat{k}$ eigenvalues of $\mathbf{L}$. Apply $k$-Means clustering on the rows of $\mathbf{V}$ to obtain cluster memberships $\hat{\mathbf{y}} \in [1, 2, \ldots, \hat{k}]^n$.
\end{enumerate}
Our proposed SC-pNA approach is outlined in Algorithm~\ref{alg:SC-pNA}. SC-pNA does not require any tuning, and has an $O(n^{3})$ complexity due to step 9, while the remaining steps have $O(n^{2})$ cost or lower. In contrast, ASC has a higher complexity of $O(Pn^{3}) = O(n^{4})$, where $P = \lfloor n/4 \rfloor$ is the recommended number of binarized similarity matrices.

\begin{algorithm}[H]
\caption{SC-pNA}
\begin{algorithmic}\label{alg:SC-pNA}
\STATE \textbf{Input:} Set of embeddings $\mathbf{X} \in \mathbb{R}^{n \times d}, \ p$.
\STATE \textbf{Output:} Cluster memberships $\hat{\mathbf{y}} \in [1, 2, \ldots, \hat{k}]^n$.
\vspace{0.3em}
\hrule
\vspace{0.4em}
\STATE 1. Compute $\mathbf{A}$ using equation \eqref{eq:Affinity}. 
\STATE 2. for $i = 1$ to $n$
\STATE 3. \quad Set $\mathbf{A}_{ii}: = 0$.
\STATE 4. \quad Form clusters $C_1^{(i)}$, $C_2^{(i)}$ using $k$-Means with $k: = 2$.
\STATE 5. \quad Identify $C_w^{(i)}$ using equation \eqref{eq:Identify_clusters}.
\STATE 6. Form $\mathbf{P}$ using equation \eqref{eq:pruning}.
\STATE 7. Form $\mathbf{W}$ using equation \eqref{eq:Symmetrize}.
\STATE 8. Form $\mathbf{L}$ using equation \eqref{eq:Laplacian}.
\STATE 9. Compute the $\mathbf{e_{gap}}$ using equation \eqref{eq:Eigengap}.
\STATE 10. Estimate $\hat{k}$ using equation \eqref{eq:Estimate_k}.
\STATE 11. Obtain $\hat{\mathbf{y}}$ using $k$-Means on the spectral embeddings in $\mathbf{V}$, corresponding to the $\hat{k}$ smallest eigenvalues of $\mathbf{L}$.
\end{algorithmic}
\end{algorithm}

\section{Experimental Setup}
We empirically compare the unsupervised SD performances of the proposed SC-pNA (with $p = 20\% $ retention), with the EER-$\Delta$, and the SOTA method of ASC. The performance of the semi-supervised CSC is also measured, which serves as an ideal baseline for unsupervised SD. For all methods, we set $k_{\max}$ to 10. We also empirically validate the robustness of SC-pNA by observing its performance over a range of retention percentages $p$ from $10$ to $50$, with increments of $5$. 

\subsection{Dataset}
For our experiments, we utilized the speech corpora from the third DIHARD speech diarization challenge (DIHARD-III)~\cite{ryant21_interspeech}. It comprises data from eleven diverse domains characterized by a variety in the number of speakers, conversation types and speech qualities. 
For each domain there exists two splits of development (dev) and evaluation (eval). This results in a total of twenty-two data splits over the eleven data domains. Both splits are used to evaluate the unsupervised SD methods. 


\subsection{SD system} 
We conducted our experiments using the SpeechBrain toolkit~\cite{speechbrain}. Our implementation is based on a modified version of the AMI recipe provided in the toolkit~\footnote{\url{https://github.com/speechbrain/speechbrain/tree/develop/recipes/AMI}}, which extracts speaker embeddings using an ECAPA-TDNN model~\cite{ecapa2020}, pre-trained on the VoxCeleb dataset~\cite{nagrani17_interspeech,chung18b_interspeech}. The extracted speaker embeddings are 192-dimensional, derived from the penultimate layer of the ECAPA-TDNN architecture. To ensure robust performance and eliminate the variability introduced by SAD, we utilized ground truth annotations. Each segment is set to 3.0 seconds, with 1.5 seconds of overlap between consecutive segments. 

\subsection{Evaluation metric}
We use the standard DER metric for evaluating the performance of the diarization system~\cite{anguera2012speaker}. DER is comprised of three key errors: \emph{missed speech}, \emph{false alarm of speech}, and \emph{speaker confusion error}. DER is quantified as the ratio of the combined duration of these three errors to the total duration.

\section{Results \& Discussion}

\subsection{Comparison with the SOTA}
The proposed SC-pNA (with $p = 20\%$ retention) and the EER-$\Delta$ approach are compared with the SOTA method of ASC. The resulting DERs achieved over DIHARD-III are shown in Table \ref{Table:Results}. 

\begin{table}[!ht]
    \centering
    \caption{Speaker Diarization performance in terms of DER (lower is better) for the semi-supervised CSC baseline along with the unsupervised SD methods of ASC, EER-$\Delta$, and SC-pNA, on DIHARD-III dataset.}
    \begin{tabular}{|l|l||l||l|l|l|}
    \hline
        Domain & Split & CSC & ASC & EER-$\Delta$ & SC-pNA \\ \hline \hline
        broadcast & dev & 2.03 & 4.09 & \textbf{2.41} & 2.98 \\ \cline{2-6}
        interview & eval & 3.58 & \textbf{3.67} & 6.82 & 4.77 \\ \hline
        \multirow{2}{*}{court} & dev & 1.82 & \textbf{2.73} & 16.68 & 6.04 \\ \cline{2-6}
        ~ & eval & 2.09 & \textbf{2.73} & 17.51 & 7.15 \\ \hline
        \multirow{2}{*}{cts} & dev & 8.28 & 9.73 & 21.40 & \textbf{8.22} \\ \cline{2-6}
        ~ & eval & 6.58 & 7.3 & 12.66 & \textbf{6.63} \\ \hline
        \multirow{2}{*}{maptask} & dev & 2.19 & 4.8 & 10.05 & \textbf{2.73} \\ \cline{2-6}
        ~ & eval & 1.78 & 7.47 & 5.25 & \textbf{0.92} \\ \hline
        \multirow{2}{*}{meeting} & dev & 16.11 & 18.6 & 26.65 & \textbf{16.79} \\ \cline{2-6}
        ~ & eval & 16.79 & \textbf{19.60} & 40.84 & 21.26 \\ \hline
        \multirow{2}{*}{socio lab} & dev & 3.38 & 4.6 & 8.33 & \textbf{3.08} \\ \cline{2-6}
        ~ & eval & 1.99 & 4.56 & 8.66 & \textbf{1.97} \\ \hline
        \multirow{2}{*}{webvideo} & dev & 33.94 & 41.57 & 35.68 & \textbf{30.68} \\ \cline{2-6}
        ~ & eval & 36.29 & 37.60 & 36.52 & \textbf{33.06} \\ \hline
        \multirow{2}{*}{restaurant} & dev & 29.08 & \textbf{30.46} & 52.78 & 37.31 \\ \cline{2-6}
        ~ & eval & 29.93 & \textbf{33.35} & 59.78 & 38.81 \\ \hline
        \multirow{2}{*}{audiobooks} & dev & 0.40 & 23.64 & 0.20 & \textbf{0.08} \\ \cline{2-6}
        ~ & eval & 0.50 & 27.53 & 0.18 & \textbf{0.09}  \\ \hline
        \multirow{2}{*}{clinical} & dev & 7.33 & 11.83 & 28.07 & \textbf{6.31} \\ \cline{2-6}
        ~ & eval & 4.34 & 10.62 & 31.23 & \textbf{3.31} \\ \hline
        \multirow{2}{*}{socio field} & dev & 7.01 & \textbf{7.31} & 23.38 & 9.21 \\ \cline{2-6}
        ~ & eval & 5.18 & 10.30 & 16.57 & \textbf{3.79} \\ \hline \hline
        \multirow{2}{*}{Overall} & dev & 10.76 & 13.26 & 21.10 & \textbf{10.70} \\ \cline{2-6}
        ~ & eval & 9.97 & 13.11 & 20.08 & \textbf{10.27} \\ \hline
    \end{tabular}
    \label{Table:Results}    
\end{table}

We observe that SC-pNA achieves the lowest DER among the unsupervised methods over fourteen of the twenty-two data splits. ASC achieved the lowest DER among the remaining seven data splits, whereas EER-$\Delta$ in general obtained higher DERs across the data splits. SC-pNA achieves lower DER than ASC in six domains, is competitive in three, and has higher DER in only two domains. Across DIHARD-III, SC-pNA clearly outperforms both ASC and EER-$\Delta$, with performance comparable with the semi-supervised CSC.   
SC-pNA even outperforms CSC on the data splits: cts (dev), maptask (eval), socio lab (dev and eval), webvideo (dev and eval), audiobooks (dev and eval), clinical (dev and eval), and socio field (eval).
Thus SC-pNA is recommended for unsupervised SD.
 

\subsection{Variations in DERs for different $p\%$ in SC-pNA}

\begin{figure}[!ht]
    \centering
    \includegraphics[width=\linewidth]{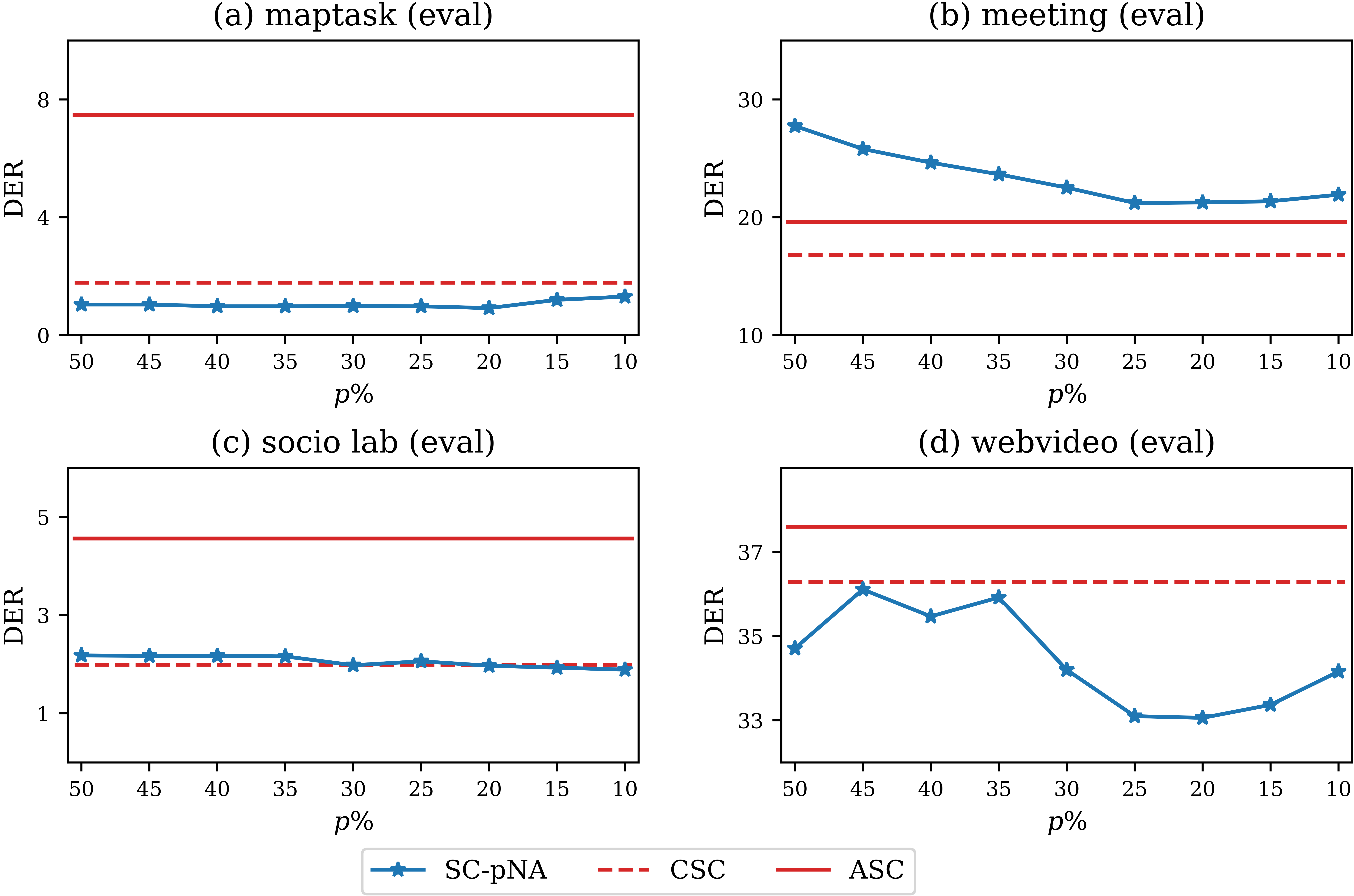}
    \caption{The figure shows the variation in the SD performance in terms of DER across four data splits, for different retention percentages obtained by SC-pNA, CSC and ASC.}
    \label{fig:vary_p}
\end{figure}

The preceding experiments evaluated the performance of SC-pNA with $p = 20\%$. Here we study the variations in DER for values of $p$ in the range of $10\%$ to $50\%$, with an increment of $5\%$. For four data splits of maptask (eval), meeting (eval), socio lab(eval), webvideo (eval), in Figure \ref{fig:vary_p} the DERs obtained by SC-pNA are compared with those obtained by CSC and ASC. Remarkably, for maptask (eval) and webvideo (eval), the DERs obtained across all retention percentages are better than both ASC and the semi-supervised CSC. For meeting (eval) the DER is generally higher than ASC or CSC, and at retention percentages of $25\%$ and $20\%$ the DERs are the closest to ASC. For socio lab (eval) the DERs are also lower than ASC, and are actually comparable to CSC. Thus we conclude that the proposed SC-pNA in general shows robust SD performance across different retention percentages, and the selected $20\%$ retention can be recommended as it generally achieves low DERs across various data splits.

\vspace{-0.5cm}
\section{Conclusion}

We proposed an SD approach called SC-pNA, which creates a sparse affinity matrix following an approach motivated by the theory on EER for Gaussian distributions. SC-pNA identifies two clusters in each row of an initial affinity matrix, and retains the top $p\%$ similarity scores from the cluster containing larger similarity scores. This approach enables SC-pNA to prune a variable number of similarity scores. The resulting sparse affinity matrix is symmetrized, from which the Laplacian is formed. The speakers are then identified by clustering the spectral embeddings, based on the number of speakers that is automatically selected from the eigengap. The overall method of SC-pNA has significantly lower computation cost compared to the SOTA ASC approach, while our empirical results show that SC-pNA with a retention percentage of $20\%$ outperforms the ASC over the DIHARD-III dataset, making SC-pNA a method to be recommended for SD. Empirical results also notably showed that SC-pNA was competitive against the semi-supervised CSC, and showed robust performances across different retention percentages.
\section*{Acknowledgment}
The primary author would like to thank his doctoral advisory committee members at RKMVERI for their feedback on the initial study on speaker diarization.
He also expresses sincere gratitude to the Linguistic Data Consortium (LDC) for awarding the LDC Data Scholarship, which enabled access to the DIHARD-III dataset.

\bibliographystyle{IEEEbib}
\bibliography{refs}

\begin{thebibliography}{10}

\bibitem{park2022review}
Tae~Jin Park, Naoyuki Kanda, Dimitrios Dimitriadis, Kyu~J Han, Shinji Watanabe, and Shrikanth Narayanan,
\newblock ``A review of speaker diarization: Recent advances with deep learning,''
\newblock {\em Computer Speech \& Language}, vol. 72, pp. 101317, 2022.

\bibitem{dawalatabad21_interspeech}
Nauman Dawalatabad, Mirco Ravanelli, François Grondin, Jenthe Thienpondt, Brecht Desplanques, and Hwidong Na,
\newblock ``{ECAPA}-{TDNN} embeddings for speaker diarization,''
\newblock in {\em Proc. INTERSPEECH}, 2021.

\bibitem{CallHome}
``{CALLHOME} {A}merican english speech,'' \url{https://catalog.ldc.upenn.edu/LDC97S42},
\newblock Accessed: 2024-09-01.

\bibitem{carletta2005ami}
Jean Carletta et~al.,
\newblock ``The {AMI} meeting corpus: A pre-announcement,''
\newblock in {\em International Workshop on Machine Learning for Multimodal Interaction}. Springer, 2005, pp. 28--39.

\bibitem{huh2023voxsrc}
Jaesung Huh, Andrew Brown, Jee-weon Jung, Joon~Son Chung, Arsha Nagrani, Daniel Garcia-Romero, and Andrew Zisserman,
\newblock ``{VoxSRC} 2022: The fourth voxceleb speaker recognition challenge,''
\newblock {\em arXiv preprint arXiv:2302.10248}, 2023.

\bibitem{ryant21_interspeech}
Neville Ryant, Prachi Singh, Venkat Krishnamohan, Rajat Varma, Kenneth Church, Christopher Cieri, Jun Du, Sriram Ganapathy, and Mark Liberman,
\newblock ``The third {DIHARD} diarization challenge,''
\newblock in {\em Proc. INTERSPEECH}, 2021.

\bibitem{jain2010data}
Anil~K Jain,
\newblock ``Data clustering: 50 years beyond k-means,''
\newblock {\em Pattern Recognition Letters}, vol. 31, no. 8, pp. 651--666, 2010.

\bibitem{ren2024deep}
Yazhou Ren, Jingyu Pu, Zhimeng Yang, Jie Xu, Guofeng Li, Xiaorong Pu, S~Yu Philip, and Lifang He,
\newblock ``Deep clustering: A comprehensive survey,''
\newblock {\em IEEE Transactions on Neural Networks and Learning Systems}, 2024.

\bibitem{salmun2017plda}
Itay Salmun et~al.,
\newblock ``{PLDA}-based mean shift speakers' short segments clustering,''
\newblock {\em Computer Speech \& Language}, vol. 45, pp. 411--436, 2017.

\bibitem{ning2006spectral}
Huazhong Ning et~al.,
\newblock ``A spectral clustering approach to speaker diarization,''
\newblock in {\em Proc. INTERSPEECH}, 2006.

\bibitem{Park2020AutoTuningSC}
Tae~Jin Park et~al.,
\newblock ``Auto-tuning spectral clustering for speaker diarization using normalized maximum eigengap,''
\newblock {\em IEEE Signal Processing Letters}, vol. 27, pp. 381--385, 2020.

\bibitem{ng2001spectral}
Andrew Ng, Michael Jordan, and Yair Weiss,
\newblock ``On spectral clustering: Analysis and an algorithm,''
\newblock in {\em Proc. Advances in Neural Information Processing Systems}, 2001, vol.~14.

\bibitem{von2007tutorial}
Ulrike Von~Luxburg,
\newblock ``A tutorial on spectral clustering,''
\newblock {\em Statistics and Computing}, vol. 17, pp. 395--416, 2007.

\bibitem{shi2000normalized}
Jianbo Shi and Jitendra Malik,
\newblock ``Normalized cuts and image segmentation,''
\newblock {\em IEEE Transactions on Pattern Analysis and Machine Intelligence}, vol. 22, no. 8, pp. 888--905, 2000.

\bibitem{poh2005f}
Norman Poh and Samy Bengio,
\newblock ``F-ratio client dependent normalisation for biometric authentication tasks,''
\newblock in {\em Proc. ICASSP}, 2005.

\bibitem{prakash07_interspeech}
Vinod Prakash and John H.~L. Hansen,
\newblock ``Score distribution scaling for speaker recognition,''
\newblock in {\em Proc. INTERSPEECH}, 2007.

\bibitem{dehak2010front}
Najim Dehak, Patrick~J Kenny, R{\'e}da Dehak, Pierre Dumouchel, and Pierre Ouellet,
\newblock ``Front-end factor analysis for speaker verification,''
\newblock {\em IEEE Transactions on Audio, Speech, and Language Processing}, vol. 19, no. 4, pp. 788--798, 2010.

\bibitem{snyder2018x}
David Snyder, Daniel Garcia-Romero, Gregory Sell, Daniel Povey, and Sanjeev Khudanpur,
\newblock ``X-vectors: Robust dnn embeddings for speaker recognition,''
\newblock in {\em Proc. ICASSP}. IEEE, 2018, pp. 5329--5333.

\bibitem{bai2021speaker}
Zhongxin Bai and Xiao-Lei Zhang,
\newblock ``Speaker recognition based on deep learning: An overview,''
\newblock {\em Neural Networks}, vol. 140, pp. 65--99, 2021.

\bibitem{shum12_interspeech}
Stephen Shum, Najim Dehak, and James Glass,
\newblock ``On the use of spectral and iterative methods for speaker diarization,''
\newblock in {\em Proc. INTERSPEECH}, 2012.

\bibitem{lin19_interspeech}
Qingjian Lin et~al.,
\newblock ``{LSTM based similarity measurement with spectral clustering for speaker diarization},''
\newblock in {\em Proc. INTERSPEECH}, 2019.

\bibitem{raghav2024assessing}
Nikhil Raghav and Md~Sahidullah,
\newblock ``Assessing the robustness of spectral clustering for deep speaker diarization,''
\newblock in {\em Proc. IEEE INDICON}, 2024.

\bibitem{brummer2014generative}
Niko Br{\"u}mmer and Daniel Garcia-Romero,
\newblock ``Generative modelling for unsupervised score calibration,''
\newblock in {\em Proc. ICASSP}, 2014.

\bibitem{speechbrain}
Mirco Ravanelli et~al.,
\newblock ``{SpeechBrain}: A general-purpose speech toolkit,''
\newblock {\em arXiv preprint arXiv:2106.04624}, 2021.

\bibitem{ecapa2020}
Brecht Desplanques et~al.,
\newblock ``{ECAPA-TDNN}: Emphasized channel attention, propagation and aggregation in tdnn based speaker verification,''
\newblock in {\em Proc. INTERSPEECH}, 2020.

\bibitem{nagrani17_interspeech}
Arsha Nagrani, Joon~Son Chung, and Andrew Zisserman,
\newblock ``{VoxCeleb}: A large-scale speaker identification dataset,''
\newblock in {\em Proc. INTERSPEECH}, 2017.

\bibitem{chung18b_interspeech}
Joon~Son Chung, Arsha Nagrani, and Andrew Zisserman,
\newblock ``{VoxCeleb2}: Deep speaker recognition,''
\newblock in {\em Proc. INTERSPEECH}, 2018.

\bibitem{anguera2012speaker}
Xavier Anguera et~al.,
\newblock ``Speaker diarization: A review of recent research,''
\newblock {\em IEEE Transactions on Audio, Speech, and Language Processing}, vol. 20, no. 2, pp. 356--370, 2012.

\end{thebibliography}

\end{document}